\documentclass[journal=jacsat,manuscript=article]{achemso}

\usepackage{chemformula} 
\usepackage[T1]{fontenc} 

\usepackage{color,soul}

\author{Sanjin Marion}
\author{Sebastian J. Davis}
\affiliation[EPFL]
{Laboratory of Nanoscale Biology, Institute of Bioengineering, School of Engineering, EPFL, 1015 Lausanne, Switzerland}
\author{Zeng-Qiang Wu}
\affiliation{State Key Laboratory of Analytical Chemistry for Life Science, School of Chemistry and Chemical Engineering, Nanjing University, Nanjing, 210093, China.}
\author{Aleksandra Radenovic}
\email{aleksandra.radenovic@epfl.ch}
\affiliation[EPFL]
{Laboratory of Nanoscale Biology, Institute of Bioengineering, School of Engineering, EPFL, 1015 Lausanne, Switzerland}

\title[]{Nanocapillary Confinement of Imidazolium Based Ionic Liquids}


\begin{document}
\begin{abstract}

Room temperature ionic liquids are salts which are molten at or around room temperature without any added solvent or solution. In bulk they exhibit glass like dependence of conductivity with temperature as well as coupling of structural and transport properties. Interfaces of ionic liquids have been found to induce structural changes with evidence of long range structural ordering on solid-liquid interfaces spanning length scales of $10-100$nm. Our aim is to characterize the influence of confinement on the structural properties of ionic liquids. We present the first conductivity measurements on ionic liquids of the imidazolium type in single conical glass nanopores with confinements as low as tens of nanometers. We probe glassy dynamics of ionic liquids in a large range of temperatures ($-20$ to $70^\circ$C) and nanopore opening sizes ($20-600$nm) in silica glass nanocapillaries. Our results indicate no long range freezing effects due to confinement in nanopores with diameters as low as $20$nm. The studied ionic liquids are found to behave as glass like liquids across the whole accessible confinement size and temperature range. 

\end{abstract}

\maketitle

\section*{Introduction}
Room temperature ionic liquids (RTIL) are salts which, due to poorly coordinated ions, exist as a liquid near room temperature. They are promising candidates for a whole new generation of electrochemical devices such as batteries and supercapacitors. RTIL are also widely used as solvents in chemical reactions and bioscience due to their electrochemical stability and adaptability potential\cite{Galinski2006, Armand2009, Fedorov2014}, and their versatility as a next generation lubricant\cite{Zhou2009}. With over $1000$ synthesized ionic liquids\cite{Chiappe2005}, and about $10^{18}$ theoretical possibilities\cite{Keskin2007}, one could tailor ionic liquids to the particular use in question. In part as a result of their sheer number, and in part owing to the plethora of interesting and complex behaviour, the understanding of ionic liquid properties is still in its infancy. To explain basic properties like density and phase behavior one needs to take into account long range interactions \cite{Fedorov2014}. These long range Coulomb and dipole-dipole interactions warrant the inclusion of non-local contributions into standard models\cite{Kobrak2010,Kjellander2016a,Gebbie2015}, complicating the understanding of these systems.

In their bulk state, RTIL can exhibit glass like properties (e.g.\ dramatic changes of conductance and viscosity with temperature)\cite{Galinski2006}, and long range liquid crystalline ordering\cite{Goossens2016}. At interfaces the shape and size of the ions play an important factor in determining local ionic liquid structure causing a competition between overcrowding and overscreening\cite{Bazant2011}. Recent experimental work using surface force apparatus (SFA) confirms the failure of standard Debye-H\"{u}ckel theories applicable for aqueous salt solutions. These experiments indicate an increase of the electrostatic screening length in concentrated electrolytes with concentration, not only in RTIL but also in standard aqueous solutions\cite{Smith2016}. It has also been postulated that, due to the correlation between ions, neutral solvent molecules (or impurities) could act as charge carriers\cite{Lee2017}. This also causes an anomalous increase in the diffusion of small neutral ions\cite{Araque2015}. Recent research indicates that there is close coupling between RTIL structure and dynamical properties\cite{Araque2015a}, with open questions about what is the charge carrier for conductance (ions, holes, or impurities)\cite{Lee2017} and how electrostatic screening behaves in such systems\cite{Feng2019}.

Nanoconfinement of RTIL exhibits an abundance of additional phenomena, owing to the interplay of sterics and long range dipole interactions. Research on RTIL in porous materials indicates several phenomena, from local changes in ionic liquid densities in confinement\cite{Matter2012,Breitsprecher2017}, evidence of a superionic state due to configurational constraints of the ions at nanometer scales\cite{Futamura2017}, to a change of the glass transition temperature with nano-porous membranes\cite{Zuo2019}. Of course porous media such as alumina membranes have some distribution of pore sizes and shape and thus do not probe precisely the exact confinement state of ionic liquids. Optical measurements on surfaces\cite{Anaredy2016,Ma2016,Pontoni2019} and small angle x-ray scattering (SAXS) on colloidal suspensions of particles\cite{Kamysbayev2019} indicate long range structural ordering mediated by surfaces. Experiments similar to atomic force microscopy using tuning forks have recently demonstrated long range structural ordering at room temperature near surfaces on length scales in the range of $\sim 10-150$nm akin to a freezing transition\cite{Comtet2017}. This transition is found to be stronger near metal surfaces due to the stabilization of the ionic liquid crystal-like lattice by induced image charges in the metal. On the other hand, SFA experiments detect only ordering of several molecular layers on insulating surfaces\cite{Lhermerout2018}, attributing this phenomena to a preweting surface layer of several nanometers. Garcia \textit{et al.}\ claim the existence of a prewetting layer which could expand up to tens of nanometers on metallic surfaces\cite{Garcia2017}. 

Clearly, current techniques for studying ordering of ionic liquids on the nanoscale can not yet give a clear picture of the underlying physics behind any structural ordering effects. Probing for the influence of single nanopore confinement on ionic liquids would provide unprecedented access to the underlying physics of correlations between ions and help resolve the open questions relating to their structure and dynamics. An early study of an ionic liquid confined in a single conical nanopore showed an abrupt drop in conductance of the pores as the confinement was decreased below $\sim 50$nm\cite{Davenport2009}, albeit without any explanation on the nature of such a change. The current work presents a study of transport properties of ionic liquids, via AC conductivity measurements with inert, non-contaminating electrodes, in single nanopore confinement as a function of temperature. These measurements are used to elucidate how confinement influences the structure of ionic liquids.

\begin{figure}[h!]
\centering
\includegraphics[width=\textwidth]{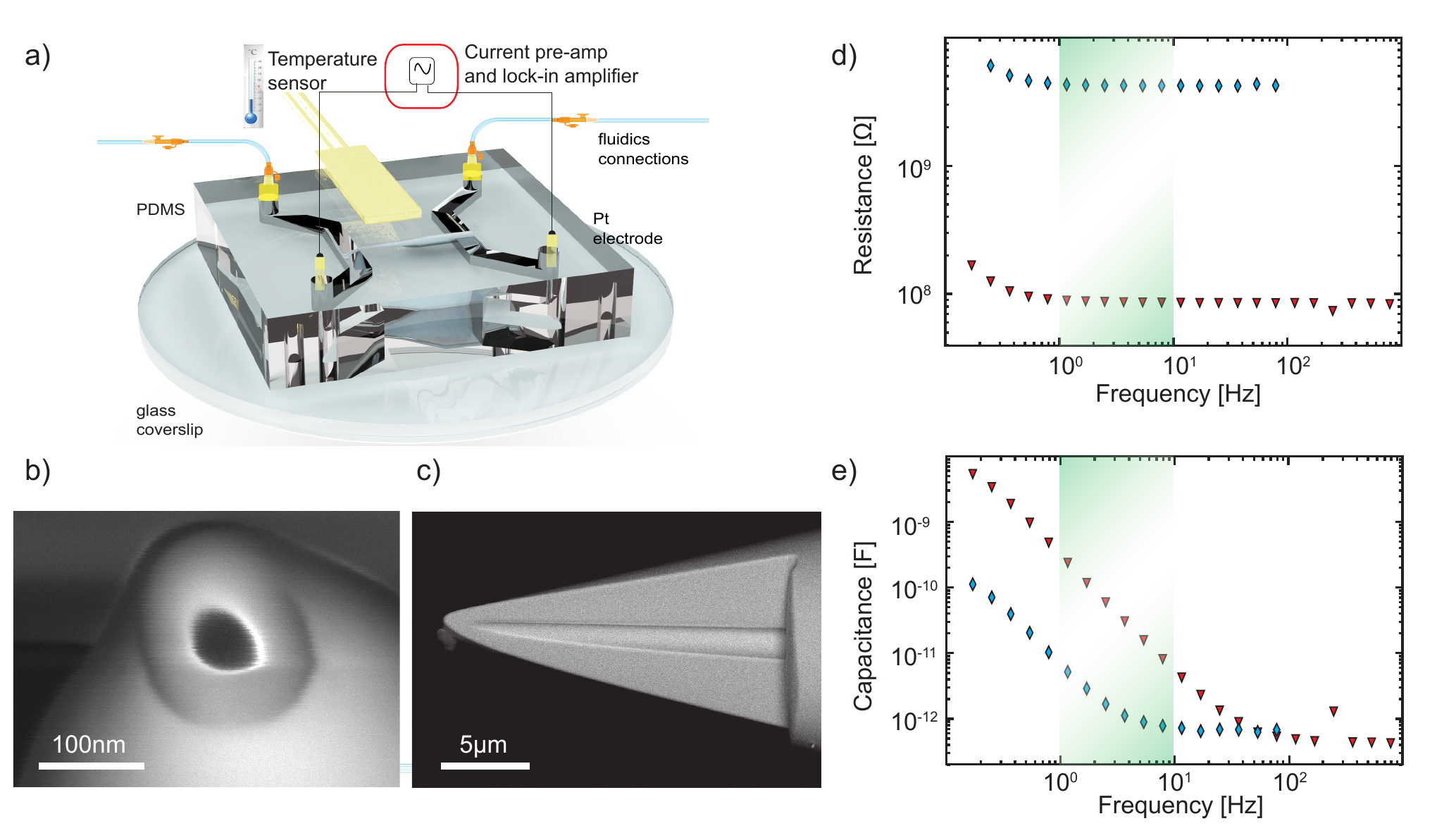}
\caption{\small \textbf{Experimental setup and frequency measurements.} a) Schematic of sample mounting, temperature and lock-in alternating current (AC) electronic measurements. A Peltier element placed below the glass coverslip is used to regulate the temperature of the sample. b) A typical SEM micrograph taken vertically above the capillary opening. This is used to measure the pore size of each capillary as well as to confirm that the opening is circular and symmetric. c) Cross section of a typical nanocapillary. This is done by focused ion beam (FIB) etching of the capillary until the inner diameter is visible. Micrographs such as these were taken in order to verify that the inner channel of typical capillaries were standard tapers. They were not used to measure the size of the opening. Note that the slice does not go completely through the middle of the opening at the tip side. d) Example of the frequency response of the resistance of a typical capillary obtained by sweeping an AC signal and assuming the sample can be modeled as a parallel connection of a resistor $R$ and a capacitor $C$. This particular case is a $50$nm capillary filled with emimBF4. The blue coloured trace is measured at $T = -22^\circ$C while the red trace is measured on the same capillary at $T = 70^\circ$C. This shows the standard behaviour expected in nanocapillaries with some lower frequency electrode polarisation before a plateau region where the measured resistance is exactly equivalent to DC conductivity measurements. In all following results one frequency within the shaded plateau region is chosen to probe the sample. e) Frequency response of the capacitance taken on the same capillary as Fig. \ref{fig:intro}d). This shows that the capacitance stabilises, after electrode polarisation effects wear off in the higher frequency regime, to an expected value of $<1$pF for the glass capillaries.}\label{fig:intro}
\end{figure}

\section*{Results and Discussion}
\begin{figure}[h!]
\centering
\includegraphics[width=\textwidth]{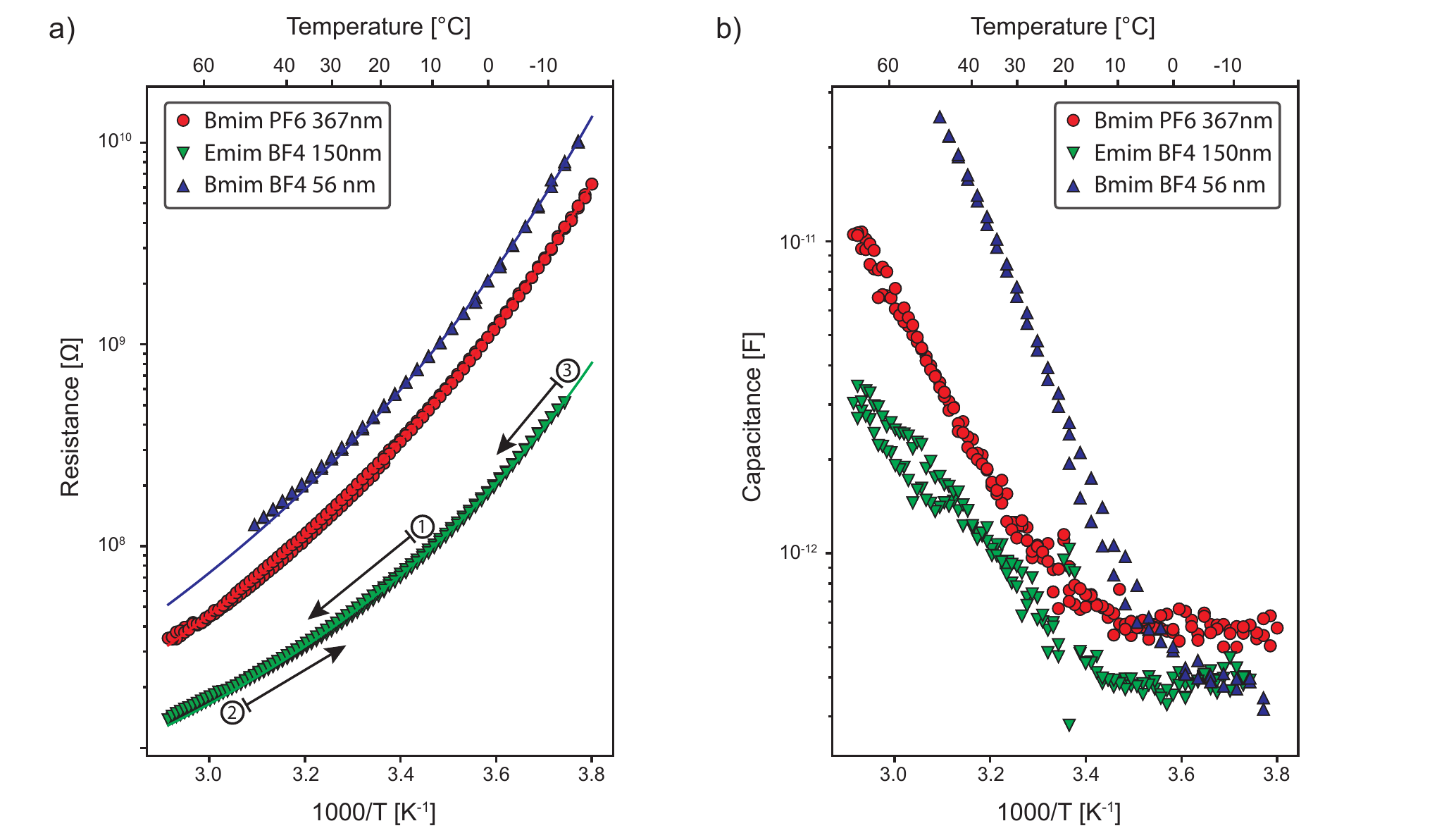}
\caption{\small \textbf{Temperature dependent resistance and capacitance of ionic liquids in single pore confinement.} Typical temperature dependent resistance (a)) and capacitance (b)) of the three ionic liquids studied. BmimBF4 (blue) is measured in a $56$nm capillary, emimBF4 (green) in a $150$nm capillary, and bmimPF6 (red) in a $367$nm capillary. These are chosen as they show the complete range of resesitances measured and that the RTIL follow the expected behaviour over this whole range. The VFT formula was fitted to each resistance curve by fixing a value of $B=-800$ for bmimBF4 and bmimPF6, and $B=-600$ for emimBF4 according to published measurements\cite{Zech2010,Stoppa2010}. It is shown as a solid line. Capacitance values saturate to values expected for glass nanocapillaries. All temperature sweeps follow the protocol described by arrows on panel a). That is to say the sample starts at room temperature, is brought to $70$ deg C, then the temperature is reduced to $-20$ deg C before being brought back to room temperature. This allows the measurement of any hysteresis in the sample.}\label{fig:temp_compare}
\end{figure}

For the first time, we probed nanoconfinement of ionic liquids in conical glass nanocapillaries and its influence on their transport properties. Glass nanocapillaries were used because of their chemical resistance and exceptionally low capacitance ($<1$pF) allowing precise measurements of highly resistive samples, up to tens of G$\Omega$. Fused silica nanocapillaries, previously pulled to opening sizes between $20-600$nm were filled (Fig.\ \ref{fig:intro}b,c), using three different ionic liquids of the imidazolium type (emimBF4, bmimBF4 and bmimPF6). The choice of ionic liquids is made following the seminal work by Comtet et al. \cite{Comtet2017} where emimBF4 is used. Ionic liquids of the same imidazolium family with slight variations of both the cation and anion are thus chosen for the following study. The transport properties in a wide temperature range ($-20$ to $70^\circ$C) are then measured by applying a sinusoidal voltage through platinum electrodes and reading the response current using a lock-in amplifier (Fig.\ \ref{fig:intro}a). From the resulting measured current and phase shift the resistance $R$ and capacitance $C$ of the system is computed assuming a parallel connection of a resistor and capacitor. As shown on Fig.\ \ref{fig:temp_compare}a) the temperature was first raised to $70^\circ$C from room temperature before lowering to $-20^\circ$C and finally returning to room temperature. This ensures that there is no change of the sample over time (ageing or contamination) as well as checks for any induced hysteresis (for example from a freezing event). The aim of this study being to probe possible freezing, pre-wetting, or ordering effects as previously observed at room temperature on surfaces or in porous membranes. Single pore measurements are important as they remove any possible effects of wide pore size distributions or irregular shapes. If freezing were to occur it is expected that the transport properties would dramatically change near the critical confinement and temperature. The wide temperature range studied here allows to probe a larger phase space in order to investigate the presence of a freezing transition under confinement.

To test the frequency response of the ionic liquids, the sinusoidal set voltage was swept between $0.1$Hz and $1$kHz (Fig.\ \ref{fig:intro}d,e). The frequency response of the current through ionic liquid filled nanocapillaries was similar to that of aqueous filled pores. Namely electrode polarization effects are visible at low frequencies as an increase in effective resistance and capacitance, but are suppressed by the high resistance of the nanopores measured at frequencies higher than $\sim 1$ Hz. As the frequency is raised a plateau where the response is dominated by the bulk (non-electrode) value is obtained before cable leakages and the amplifier bandwidth start to dominate ($>100-1000$ Hz). This behaviour is equivalent to what is seen in nanocapillaries filled with aqueus solutions, such as KCl, where the resistance value of the plateau corresponds exactly to the DC one as obtained from a current voltage curve. This confirms that the measurements of both resistance and capacitance, as taken from the plateau region, are accurate and concurrent with our understanding of the system\cite{Marion19}. No conductivity relaxations were found in the studied range of frequency at any studied temperature. This is consistent with literature where a relaxation is expected but is present at much higher frequencies ($\sim$MHz) in the temperature range studied here\cite{Matter2012}. To eliminate any artefacts from the electrodes or the instrumentation used, all subsequent measurements were done at one fixed frequency between $1-10$ Hz (shaded region Fig. \ref{fig:intro}d,e) for each sample. This enables precise measurement of the resistance of the nanopore without having to take into account any electrochemical effects on the electrodes which plague DC measurements. 

If the hypothesis that ionic liquids have a confinement induced phase transition\cite{Comtet2017} is correct, we expect an abrupt change in the transport properties of the ionic liquids to match the abrupt increase of viscosity in the pore vicinity. This is corroborated by a study of two imidazolium based ionic liquids in chemically modified polymer track etched nanopores which shows an abrupt drop in conductance as the diameter of the nanopores is reduced below $\sim 50$nm\cite{Davenport2009}. Alternatively, if there is an ordered (frozen) layer near the surface, the resistance is expected to behave as if the nanopore diameter is effectively reduced in size. The transition temperature of such an effect is expected to follow the so called Gibbs-Thompson relation\cite{Alba-Simionesco2006,Comtet2017} which takes into account the influence of the surface energy of the solid-liquid interface in shifting the phase transition in temperature. According to the Gibbs-Thompson relation a change of the freezing transition temperature $\Delta T$ from the bulk value $T_B$ should be inversely proportional to the degree of confinement: $\Delta T \approx T_B \Delta\gamma / \rho L_h D $, with $\Delta \gamma$ the difference of surface energies between a liquid and solid phase in respect to the membrane-ionic liquid interface, $L_h$ the latent heat of melting, $\rho$ the density of the liquid phase, and $D$ the diameter of confinement. Using literature values for bmimBF4 \cite{Comtet2017}, one would anticipate that a freezing transition should be detectable in a range of nanopores close to $25$nm in diameter and with a temperature range of $70$ to $-20^\circ$C. This is all the more true of emimBF4 due to its larger native glass transition temperature (the melting point for emimBF4 is $12^\circ$C\cite{Fuller97} while its glass temperature is $-124^\circ$C\cite{Vila07}). We assume that if an ionic liquid would freeze in a certain size of nanopore it will also not fill the pore completely. To ensure complete wetting of the pores prior to temperature sweeps, in particular lower temperatures since they will more likely induce ordering effects, the capillary tip is heated up to $80^\circ$C during the filling procedure. 

Nanocapillaries filled with ionic liquids show a large increase in the resistance of the sample as the temperature is decreased due to a slowdown of charge carrier dynamics (glass-like behavior). A modified Walden rule conveys that the resistivity of an ionic liquid $\sigma$ is proportional to its viscosity $\sigma \sim \eta^\alpha$, with $\alpha\approx 0.9$ shown to be valid for emimBF4 and bmimBF4\cite{Schreiner2010}. Thus by measuring the change in conductance of the sample information about a relative change of the local viscosity in the nanocapillary tip is obtained. Fig.\ \ref{fig:temp_compare} demonstrates typical measured resistances and capacitances versus temperature for three different ionic liquids. Depending on the opening diameter of the nanocapillary the measured resistance values varied over more than three orders of magnitude.
The measured effective capacitance (Fig.\ \ref{fig:temp_compare}b)) is extremely sensitive to the working frequency due to the high resistance of the sample. As the temperature of the sample is reduced, electrode polarization effects become less pronounced as they are overshadowed by the nanopore resistance. As we reach low temperatures, the capacitance value stabilizes to the value of $0.3-0.7$pF, comparable to capacitance values obtained for glass capillaries using aqueous salt solutions. The temperature dependence of the resistance closely follows the Vogel-Fulcher-Tammann (VFT) empirical formula for charge transport in glass like systems\cite{Angell1995}:
\begin{equation}
R(T) = R_0 e^{- B / (T-T_0)}
\label{eq:VFT}
\end{equation}
where $R_0$ is an effective resistance, $B$ a parameter linked to an activation barrier for charge transport, and $T_0$ an effective glass transition temperature. The values obtained for $B$ and $T_0$ are consistent with the literature values of $B=-607.5$ ($B=-806.6$) and $T_0=163.7$ ($T_0=167$) for emimBF4 (bmimBF4)\cite{Stoppa2010}. We note that the large error in our case comes from difficulties in filling single pores with ionic liquids without contamination. This confirms that, at least in larger single nanopores, the studied ionic liquids still behave as glass-like liquids as the temperature is decreased.

\begin{figure}[h!]
\centering
\includegraphics[width=\textwidth]{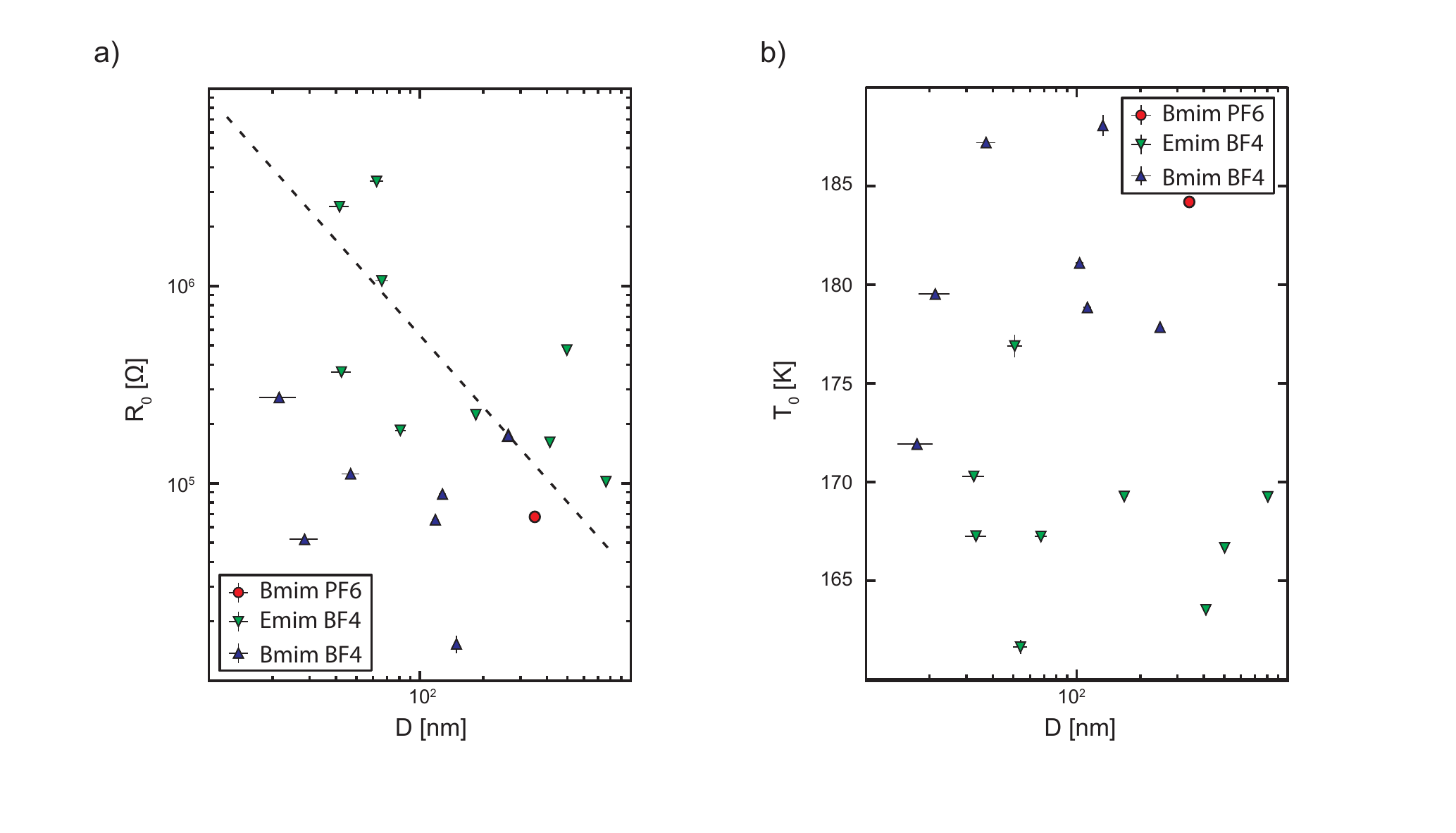}
\caption{\small \textbf{Vogel-Fulcher-Tammann fit parameters for different ionic liquids in single capillary confinement.} a) The effective resistance $R_0$ extracted from the VFT fit of temperature dependent measurements as a function of capillary pore size. The dashed line is eq. (\ref{eq:cap_res}) to show standard resistance versus tip diameter scaling for nanocapillaries. b) $T_0$ parameter extracted from VFT fits as in (a). The error bars for both $R_0$ and $T_0$ are taken from the VFT fit. The error bars on nanocapillary pore sizes are arbitrarily fixed at $\pm 5nm$ to take into account any drift or charging due to imaging a dielectric in SEM as well as to allow for any small irregularities in the shape of the pore.}\label{fig:VFT_compare}
\end{figure}

Measurements of the influence of confinement on the ionic liquid transport properties are shown in Fig.\ \ref{fig:VFT_compare}. The diameter of the nanocapillary opening $D$ was varied and a full temperature dependent sweep as in Fig.\ \ref{fig:temp_compare} was performed. The parameters $R_0$ and $T_0$ are then extracted from a fit to the VFT equation \eqref{eq:VFT}. Results indicate no significant change in behavior due to confinement down to $20$nm, apart from a trend where the resistance increases as the nanocapillary opening size $D$ is decreased. This dependence of $R_0$ on $D$ matches the expected scaling for nanocapillaries as determined in aqueous salt solutions\cite{Steinbock2013}. The dashed line shown on Fig.\ \ref{fig:VFT_compare}a corresponds to the expected dependence of nanocapillary resistance on the opening diameter $D$ 
\begin{equation}
R = \sigma^{-1} \left(\frac{4t}{\pi D d} + \frac{1}{2D} + \frac{1}{2d}\right)
\label{eq:cap_res}
\end{equation}
with $d$ the diameter of the capillary at the wide end and $t$ the taper length. The effective resistance at the glass temperature $R_0$ is higher for emimBF4 than for bmimBF4 and can be explained by the different slopes of the resistance versus temperature curves. Interestingly we do not observe any dependence of the effective glass temperature $T_0$ on the pore size $D$ within the experimental error. In addition no abrupt changes in the resistance in the available temperature range are observed for any of the ionic liquids. This can be explained as there being no freezing transition and no influence of the surface on the ordering inside the capillary, within the experimental error, in contrast to ref\cite{Comtet2017}. Another possibility is that, as compared to a recent study where a change in glass temperature was observed\cite{Zuo2019}, there is an issue with the purity of the sample which affects the ionic liquids' ability to freeze. It is also possible that the ordering effect is too small with our combinations of ionic liquids and surface material (amorphous glass). Due to the high viscosity of the ionic liquid samples only a modest percentage of attempted nanocapillaries were successfully filled, estimated at 30\% limiting our statistics.

The mechanism of conductance of ionic liquids is known to be highly sensitive to the types and quantities of impurities\cite{Zech2010a,Stoppa2010}, out of which ambient water absorption can strongly influence the bulk conductivity\cite{Widegren2005a,Widegren2005} and possibly even modify structural properties\cite{Ma2018,Liu2017,Seddon2009}. In the present case, filling of the nanocapillaries in a completely controlled atmosphere, such as a nitrogen glove box, was not possible due to their fragility and problems with handling. Another possibility is the presence of air bubbles in the sub-micron range which would not be visible using an optical microscope\cite{Marion19}. These bubbles could be partially obstructing the pore at the tip or modifying the surface of the taper thus causing a large variance in the resistance versus diameter dependence of the nanocapillaries. They could also be influencing the ordering of the RTIL itself. Surface nanobubbles have been reported in viscous media\cite{An2015}, but in ionic liquids there have so far only been optical measurements proving micron sized and larger bubbles\cite{Taylor2017}. Another possibility for the large variance in the resistance versus diameter is that nanocapillaries have by themselves an ill defined geometry due to the production method, and that small variations of the tapering angle are not controllable. In order to increase any ordering effects platinum coated nanocapillaries were also used (analogously to ref. \ \citenum{Comtet2017}). Unfortunately in this configuration the addition of a platinum layer had the effect of the current short-circuiting through the platinum surface coating inside the nanopore and bypassing the area of largest resistance (and confinement). We conclude that a different approach is needed to probe the influence of metallic surfaces on ionic liquid ordering.

\section*{Conclusion}

We have demonstrated measurements of temperature dependent transport properties of ionic liquids of the imidazolium type confined in a single nanopore using glass nanocapillaries. Our data indicates that the ionic liquids under nanoconfinement in amorphous glass nanopores behave as glassy liquids, analogously to their reported bulk behavior. No change in their glass like properties, nor any abrupt freezing transitions are seen as the temperature or the diameter of the nanopore is varied. We conclude that, within our experimental parameters, the degree of nanoconfinement in single pores down to 20 nm induces no structural, and thus dynamical changes in the studied ionic liquids. We speculate that issues with sample purity and the variability of the nanocapillary geometries could be reducing any structural ordering in the ionic liquids and causing the large observed variability of the parameters. We propose that further studies on ionic liquids in single pore nanoconfinement should be attempted using better defined geometries (e.g. nanopores in SiN membranes) and under controlled atmospheric conditions to preserve the purity of the samples. Ionic liquids with larger side chain lengths could show enhanced ordering behavior due to a larger ionic dipole moment, e.g.\ with 1-methyl-3-octyl-imidazolium tetrafluoro-borate\cite{Zuo2019}. Similar long chain ionic liquids have been demonstrated to present surface liquid crystal phases\cite{Pontoni2019}. Studying in detail, and in single pore confinement, the exact nature of the effect of impurities would also be of great interest. Expanding the array of tools available to study RTIL behaviour on the nansocale could help the understanding of their physical and chemical properties but also enable studying effects of confinement on single molecular chemical reactions (e.g. bond forming and breaking)\cite{Lin2018a}.

\section*{Materials and Methods}
Fused silica capillaries with an outer diameter of $360\mu$m and inner diameter of $50\mu$m (BGB) were pulled to nanopores in the range of $20-1000$nm using a P-2000 laser-assisted puller (Sutter Instrument) with a program containing 2 lines (Line 1: heat 540, filament 4, velocity 10, delay 145, Line 2: heat 540, filament 4, velocity 10, delay 145, pull 160). All pulled capillaries were imaged using a scanning electron microscope to measure the pore diameter and verify the tip shape. The beam parameters for the SEM were $3$kV accelerating voltage, $350$pA beam current as in Fig.\ \ref{fig:intro}b). For the images taken after FIB slicing (Fig.\ \ref{fig:intro}c)) the SEM beam parameters were $5$kV accelerating voltage and $944$pA beam current. Only nanocapillaries with a circular, symmetric opening were used in this work.

Ionic liquids emimBF4 (1-ethyl-3-methylimidazolium tetrafluoroborate), bmimBF4 (1-butyl-3-methylimidazolium tetrafluoroborate) and bmimPF6 (1-butyl-3-methylimidazolium hexafluorophosphate) were purchased from Sigma-Aldrich (respectively $>99\%$ purity and $<1000$ppm $H_2O$, $>97\%$ purity and $<1\%$ $H_2O$, $>97\%$ purity). All ionic liquids were filtered through a $20$nm filter (Anotop 25 plus) inside a nitrogen dry box to minimize ambient water contamination as well as any large contaminants. The capillaries were first oxygen plasma cleaned for $10$ minutes before being pre-filled with ionic liquid by immersing them in a drop and desiccating for $10$ minutes. They were then mounted, using microfluidic connectors, to the tip of a syringe. The capillaries were then filled by applying pressure using a syringe pump while the tips of the capillaries were immersed in a vial of ionic liquid heated to $80^\circ$C. The capillaries were left to fill overnight. After filling, all capillaries were visually inspected using an optical microscope to confirm no particulates or air bubbles were present in the tip. Viable capillaries were then embedded in a PDMS fluidic cell consisting of two chambers as in Ref.\ \citenum{Bulushev2015} (Fig.\ \ref{fig:intro}a).

After these filling and mounting steps the samples were bonded to a Peltier element using thermal paste, such that only a $100\mu$m glass slide was dividing the ionic liquid chamber and the Peltier element. A $100\Omega$ platinum resistor was glued on top of the sample, between the two reservoirs, using thermal epoxy. This sample was enclosed in a chamber in which $99.9$\% pure nitrogen over-pressure was applied during all measurements. The temperature of the sample was controlled using a home-made PID controller written in LabVIEW. Temperature stability was within $0.05$K, with a workable temperature range from $-20$ to $70^\circ$C. After a change of temperature a $2$ minute wait time was applied after which the temperature stabilized to within $0.05$K of the target temperature. All temperature curves of the resistance and capacitance were done by cycling: first increasing the temperature from room to the maximum temperature, then decreasing to the minimal working temperature, and again going back to room temperature. Fitting was done to the full curves consisting of two points at the same temperature, and showed practically no ageing effects, increased contamination from the ambient atmosphere, or changes due to plasma cleaning, during a typical measurement ($\sim 6 h$).

All electrical measurements were done using a SR865A Lock-in amplifier (Stanford Research Systems) using either the DLPCA-200 current to voltage converter (FEMTO Messtechnik GmbH) or a ADA4530-1 evaluation board with a fixed $10$nA/V gain. The ADA4530-1 current preamplifier was tested to have less than $10$fA input bias current using standard resistors. All measurements were done with an AC bias in the frequency range of $0.1$Hz to $1$kHz ($100$Hz for ADA4530-1), depending on the amplifier's working range. A sinusidal set voltage of RMS amplitude $100$mV was applied through platinum electrodes ($90\%$ Pt $10\%$ Ir wire, Goodfellow). The amplitude $I$ and phase delay $\theta$ of the resulting current through the nanocapillary were then measured. In the range of frequencies used, we found that an effective circuit of a parallel resistor $R$ and capacitance $C$ completely explains the measured signal. The resistor value $R$ is connected to the diameter and general shape of the capillary through eq.\ \eqref{eq:cap_res}\cite{Willmott2011a,Kowalczyk2011,Steinbock2013}. All components, connectors, and wires, including the PDMS chamber and glass slides were cleaned by sonicating in isopropanol for $15$ minutes and baking for $15$ minutes at $100$ deg C before use. Several tests of the leakage current of the sample chamber were done with a capillary with a closed tip and with $1$M KCl aqueous solution in the surrounding chambers. Using the ADA4530 transconductance amplifier the sample showed a resistance of $\sim 10$T$\Omega$ between the two chambers, defining the precision of the measurement set-up.

\section{Author Contributions}
S.M. and A.R. designed the study. S.M. performed the experiments. S.J.D and Z.-Q. W. prepared the samples. S.M. and S.J.D. wrote the paper. All authors provided important suggestions for the experiments, discussed the results, and contributed to the manuscript.

\begin{acknowledgement}
The authors would like to thank M. Macha for assistance in designing the experimental setup, L. Navratilova from CIME (interdisciplinary center for electron microscopy) for the FIB milling of nanocapillaries, and J. Gundlach and D. Vasilyev for useful discussions. This work was financially supported by the Swiss National Science Foundation (SNSF) Consolidator grant (BIONICBSCGI0\_157802). Z.Q. Wu was funded by the China Scholarship Council and the National Natural Science Foundation of China (grant number: 21775066).
\end{acknowledgement}

\bibliography{extracted_references.bib}

\end{document}